%% file: ms.tex
\documentclass[10pt]{article}

\pdfoutput=1 
\usepackage[pagenumbers]{cvpr} 
\usepackage{times}
\usepackage{epsfig}
\usepackage{graphicx}
\usepackage{amsmath}
\usepackage{amssymb}
 \usepackage{booktabs}
 \usepackage{multirow}
 \usepackage{soul}
 \usepackage[table,xcdraw]{xcolor}

\begin{document}

\title{Face Recognition: to Deploy or not to Deploy? A Framework for Assessing the Proportional Use of Face Recognition Systems in Real-World Scenarios}

\author{Pablo Negri\\
Instituto de Investigacion \\
en Ciencias de la Computacion (ICC),\\
UBA-CONICET.\\
Universidad de Buenos Aires, FCEyN\\
Computer Department, \\
Buenos Aires, Argentine\\
{\tt\small pnegri@dc.uba.ar}
\and
Isabelle Hupont\thanks{The views expressed in this scientific publication are purely those of the authors and may not in any circumstances be regarded as stating an official position of the European Commission.}
\\
European Commission,\\
Joint Research Centre \\
 Sevilla,Spain\\
{\tt\small isabelle.hupont-torres@ec.europa.eu}
\and
Emilia  Gomez$^*$\\
European Commission,\\
Joint Research Centre \\
 Sevilla,Spain\\
{\tt\small emilia.gomez-gutierrez@ec.europa.eu}
}


\maketitle
\thispagestyle{empty}

\begin{abstract}
Face recognition (FR) has reached a high technical maturity. 
However, its use needs to be carefully assessed from an ethical perspective, especially in sensitive scenarios. 
This is precisely the focus of this paper: the use of FR for the identification of specific subjects in moderately to densely crowded spaces (e.g. public spaces, sports stadiums, train stations) and law enforcement scenarios. 
In particular, there is a need to consider the trade-off between the need to protect privacy and fundamental rights of citizens as well as their safety.
Recent Artificial Intelligence (AI) policies, notably the European AI Act, propose that such FR interventions should be proportionate and deployed only when strictly necessary. 
Nevertheless, concrete guidelines on how to address the concept of proportional FR intervention are lacking to date. 
This paper proposes a framework to contribute to assessing whether an FR intervention is proportionate or not for a given context of use in the above mentioned scenarios. 
It also identifies the main quantitative and qualitative variables relevant to the FR intervention decision (e.g. number of people in the scene, level of harm that the person(s) in search could perpetrate, consequences to individual rights and freedoms) and propose a 2D graphical model making it possible to balance these variables in terms of ethical cost vs security gain. 
Finally, different FR scenarios inspired by real-world deployments validate the proposed model. 
The framework is conceived as a simple support tool for decision makers when confronted with the deployment of an FR system. 

\end{abstract}

\section{Introduction}\label{sec:intro}
\input{1-Introduction.tex}

\section{Background}

\input{2-Background.tex}

\section{Intervention models from other fields}\label{sec:other_models}
\input{3-2dplanes.tex}

\section{Proposed framework}
\input{4-Framework.tex}
\section{The framework in practice} \label{sec:practice}
\input{5-SampleScenarios.tex}
\section{Compliance with International Regulations}
\input{6-AI-ACT}

\section{Conclusions and future work}
\input{7-Conclusions.tex}

\end{document}

%% file: 1-Introduction.tex
Face recognition (FR) is a flexible biometric technology capable of identifying people at a distance, even without the active cooperation of the captured subjects. In the last decade, FR systems have been used for many different purposes, such as access control~\cite{lee2020face}, border control~\cite{carlos2018facial}, device/machine unlocking~\cite{wang2019design}, control of attendance~\cite{sutabri2019automatic}, missing people identification~\cite{Negri2021} and face tagging~\cite{balakrishnan2015autotag}. 

This paper focuses on the most technically advanced, albeit ethically controversial, FR application: its real-time use to identify specific subjects in moderately to densely crowded spaces (e.g. public open spaces, sports stadiums, train stations, airports, malls) and for law enforcement purposes. 
Such FR scenarios typically make use of a multi-camera system to identify people who represent a potential threat (e.g., thieves, criminals, terrorists on police records) or are being searched (e.g. missing people or kidnapped people) over multiple video streams~\cite{barquero2020long}. Software products conceived for this specific purpose are widespread on the market~\cite{hupont2022landscape}, and they are deployed by polices and law enforcement agencies worldwide. 

From a technical perspective, FR nowadays performs successfully even in highly uncontrolled situations with tens -to- hundreds of individuals in the scene, changing lighting conditions and low face resolutions. State-of-the-art face identification algorithms achieve accuracy metrics above 95\% with a false acceptance rate of $10^{-4}$ in these contexts~\cite{liu2022controllable, boutros2022elasticface}.  
Although demographic fairness (e.g. race and gender biases) in FR is still an open research area~\cite{hupont2019demogpairs,robinson2020face}, some mitigation measures are being developed, and the FR community is raising awareness of this matter and encouraging its research~\cite{grother2019face}.

While algorithmic robustness and fairness are undoubtedly key requirements for the development of FR systems, critical ethical aspects related to deployment phases have been widely under-considered.
Even assuming that an FR system is almost perfectly accurate, fair and deployed by authorities for the exclusive purpose of improving public security, its use inevitably involves an invasion of privacy as the faces of all the subjects passing by a designated area are processed to search for a potential match with a person on a watchlist. 
In this scenario, the captured subjects might not wish to be under FR surveillance and might not be aware of the system operation. 
Other rights may also be affected when FR is used in this context, such as {\it the right to freedom of expression, peaceful assembly, and association, as well as freedom of movement}, according to~\cite{UN}. 
The authority in charge of the system deployment should therefore establish the most strictest privacy-preserving mechanisms and carefully assess the use of these technologies considering the trace-off between \textit{security and privacy (or, more broadly, fundamental rights)}.

Recent Artificial Intelligence (AI) policies addressing FR have acknowledged the importance of this trade-off. 
The European AI Act proposal~\cite{AIact}
 mandates a \textit{proportionate} and \textit{strictly necessary} use of \textit{real-time remote biometric identification systems in publicly accessible spaces for the purpose of law enforcement} and requires that their deployment shall be subject to prior authorisation by a competent authority. 
The World Economic Forum has also 
 called for \textit{responsible limits on facial recognition}~\cite{louradour2021policy} in \textit{law enforcement investigations}, highlighting its \textit{necessary and proportional use}. 

\begin{figure}
\centering
  \includegraphics[width=0.99\columnwidth]{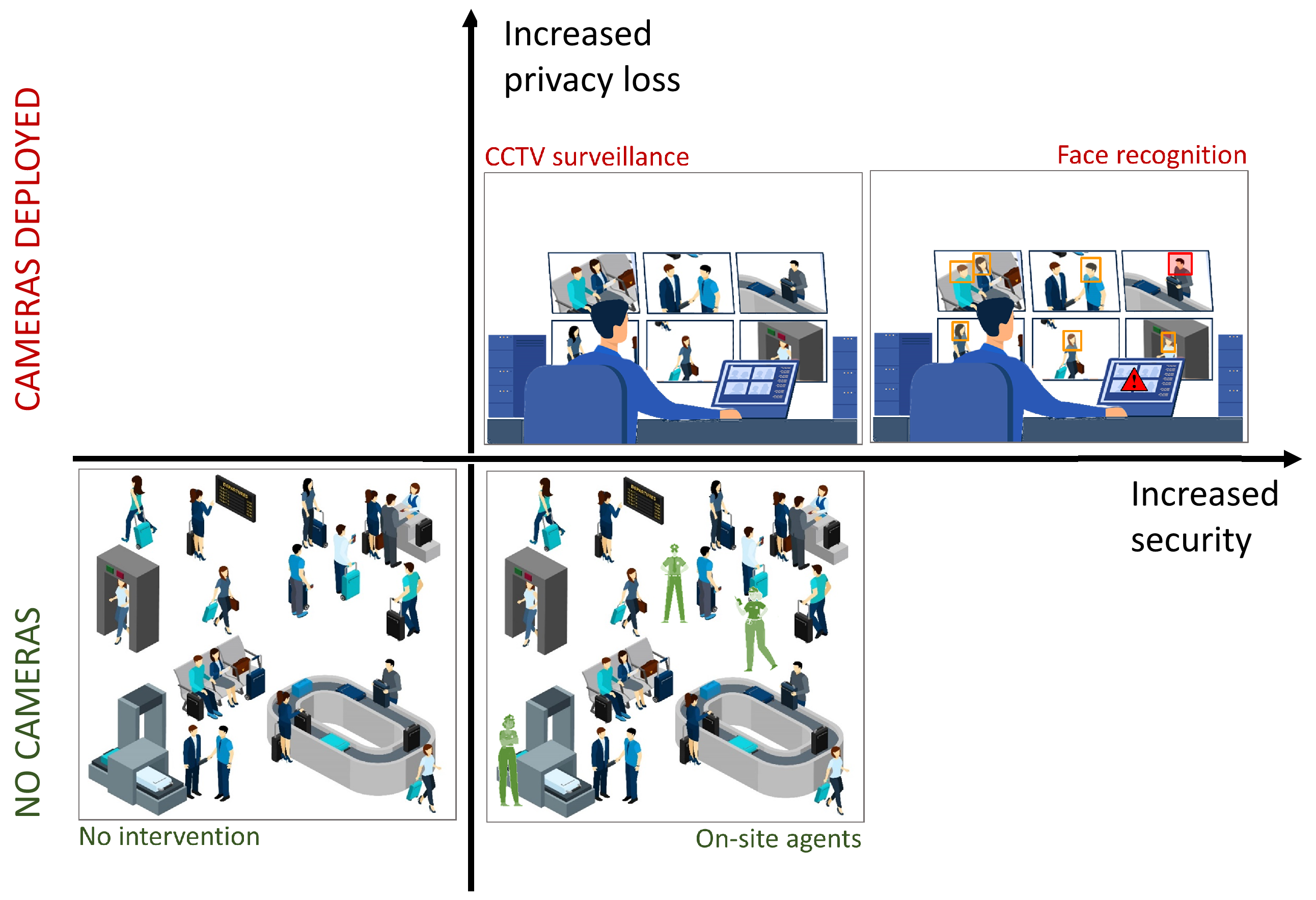}
  \caption{Different types of interventions that can be considered for a law enforcement scenario (Source: pictures modified from \cite{figmonitor, figairport}). }
  \label{fig:intervention_levels}
\end{figure}

Figure~\ref{fig:intervention_levels} illustrates four intervention alternatives that could be considered by authorities  when confronted with a \textit{security vs privacy} trade-off. \textit{No intervention} might be the best suited decision in contexts with no or very limited security needs. \textit{On-site agent intervention}, i.e., placing police agents to patrol the site, could be an alternative when security needs are higher and it is not possible to deploying cameras. 
\textit{CCTV surveillance} can be a suitable solution if cameras are available on site and real-time human supervision of streaming videos is deemed sufficient to ensure the required level of security. 
Finally, \textit{face recognition intervention} would additionally make use of an FR system to automatically analyze videos in search of faces on a watchlist and send identification alarms to security bodies.
This is the most privacy-invasive solution, although it might be needed in case of severe security threats. 

To the best of our knowledge, thus for no concrete guidelines on how to address the concept of \textit{proportional use} in FR deployments have been developed. 
Authorities would benefit from them to formalize, visualize and guide their decision on whether the deployment of an FR system 
 is proportionate or not in a given situation. 
This paper proposes a 2D framework for this assessment. 
First, the main quantitative and qualitative variables relevant to the FR deployment decision (e.g., number of people in the scene, scale of the threat, consequences on individual rights and freedoms) are identified. 
Then, a 2D model making it possible to weight these variables in terms of ethical cost (including privacy and related fundamental rights) vs security gain is proposed. 
The framework is designed to support decision makers confronted with the choice of deploying FR or not. 
Finally, the model is applied different face recognition scenarios inspired by real-world deployments, for the purposes of simulation and validation of the proposed framework.

%% file: 2-Background.tex
\subsection{The citizen perspective}\label{sec:citizen}

People make use of face recognition in their everyday life. For instance, FR is commonly used to unlock devices such as smartphones, access e-bank accounts, pass controls at airports or tag friend in social networks. 
However, when it comes to scenarios that are not so well-known or used on a daily basis, including the large-scale use of FR by law enforcement authorities for security purposes, which is the focus of this paper, studies reveal deeper reluctance to use and mistrust.

Seng et al.~\cite{seng2021first} recently analyzed people's perception of FR in 35 different scenarios, ranging from device unlocking to financial transactions, personalized marketing, control of attendance and surveillance at public events.
They showed 35 FR scenarios to 314 participants in the form of vignettes, and asked questions related to usefulness, comfort level and privacy concerns. 
Their results confirm that perceptions of FR are strongly dependent on the specific context in which it is applied. 
Participants feel more comfortable in scenarios where they trust the entities collecting their facial information and where this information is stored in their personal devices, which gives them a sense of control over their sensitive data. 
Another key finding of this study, also raised in~\cite{hupont2022landscape}, is that users who do not find a clear benefit in the use of FR in a given scenario tend to consider the technology as an invasion of privacy.
Indeed, from the 35 scenarios, only two of them relate to the large-scale use of FR at public events. They differ in one aspect: while the objective of the first one is left open (``\textit{FR is used to track people attending a public event}''), the second one specifies that the purpose is ``\textit{public safety and law enforcement}''. 
Participants found the second scenario to be more useful and reported feeling more comfortable compared to the one that did not state the purpose behind FR surveillance.
This is in accordance with \textit{the social contract theory}~\cite{moore2015privacy}, which states that individual privacy often needs to be sacrificed for the greater good such as national security. 

In addition to the benefit perceived in the use of FR for public safety purposes, recent studies have analyzed citizen trust on law enforcement agencies as the entities behind FR deployments. 
A survey with 4,109 adults run by the Ada Lovelace Institute~\cite{AdaLovelace2019}, and another one with 2,291 participants by the Monash University~\cite{Monash2020} showed that, although people have certain fears and there is no unconditional support for police use, they are open to the use of the technology for law enforcement purposes as long as there is a demonstrable public benefit, as well as regulations and privacy safeguards in the management of biometric data. 
Nevertheless, the public perception of the use of FR for law enforcement purposes is found to be closely related to cultural background. 
A study on public attitudes towards face identification in criminal justice in the USA, China, United Kingdom and Australia~\cite{ritchie2021public} found that US respondents are more accepting of citizen tracking, even though they are less trusting of the police than people in the UK and Australia. 
This illustrates the need to take into account the cultural perspective in decisions related to FR deployment by public bodies. 
The intertwining of culture and education is also important. 
As highlighted in~\cite{hupont2022landscape} relevant aspects such as the lack of knowledge of these systems by citizens (e.g., their working principles, limitations and applications) are closely connected to acceptance of this technology.

\subsection{The policy perspective}\label{sec:policy}

\begin{figure}
\centering
  \includegraphics[width=0.99\columnwidth]{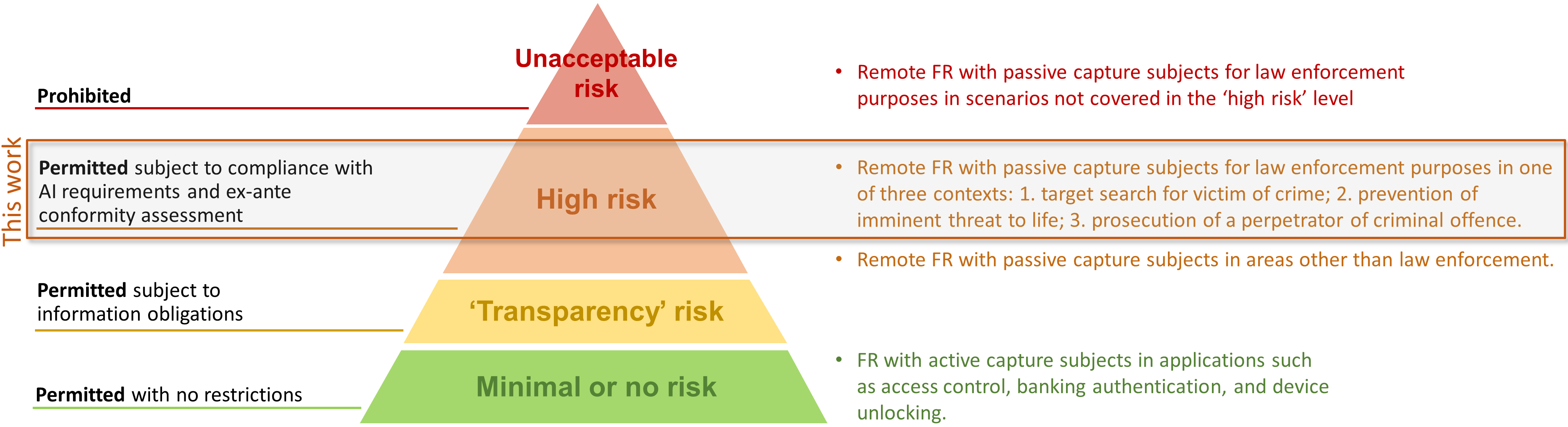}
  \caption{Illustration of the four risk levels proposed by the AI Act with the corresponding applications of face recognition. This paper focuses on the high-risk use of FR for law enforcement purposes (highlighted with an orange box). }
  \label{fig:risk_pyramid}
\end{figure}

Recent global policies refer to the \textit{proportional} use of FR technologies. 
This section considers two relevant examples. 
The first one is the European AI Act~\cite{AIact} proposal, aiming at trustworthy and safe development, implementation and use of AI systems. 
The AI Act adopts a risk based approach where AI systems are subject to different requirements according to their risk level, which is linked to their context of use and depends on how the system may impact fundamental rights. 
When this paper was written, the proposal was being refined by the European co-legislators, therefore there could be modifications to the following summary. 
We refer here to the European Commission's proposal published in 2021, which defines four risk levels: (1) Prohibited or unacceptable risk; (2) High-risk, where AI systems are subject to a set of requirements including, for example, the implementation of risk mitigation measures, appropriate levels of accuracy, robustness, cybersecurity, data governance, technical documentation and human oversight strategies; (3) Transparency risk, implying only information obligations; (4) Minimal risk, where AI systems are permitted with no restrictions. 
Hupont et al.~\cite{hupont2022landscape} analyze the landscape of facial processing applications, linking them to different risk levels like in the AI Act proposal. 
In terms of FR (Figure~\ref{fig:risk_pyramid}), the study identifies as low risk applications those intended to verify a person's identity provided that the subject has an \emph{active} role. This includes applications for access control, banking authentication or device unlocking. 
At the other side of the risk dimension, the study considers FR scenarios where subjects have a \emph{passive} role (referred to as \emph{remote} scenarios), which are linked to high or unacceptable risk based on the context. 

The proposal pays special attention to \textit{real-time remote biometric identification systems in publicly accessible spaces for the purpose of law enforcement}, which would include the real-time use of FR for law enforcement purposes studied in this paper. Article 5 explicitly regulates the use of FR for law enforcement purposes in points 1 to 3. Article 5(1d) states that this use is only allowed as far as it is \textit{strictly necessary} for one of the following three objectives: \textit{1. the targeted search for potential victims of crime; 2. the prevention of a specific, substantial and imminent threat to life or terrorist attack; and 3. the localisation, identification or prosecution of a perpetrator or suspect of a criminal offence}. The AI Act introduces security as an aspect to consider for the proportional use of FR, as Article 5(2) further specifies that this deployment shall take into account \textit{the seriousness, probability and scale of the harm caused in the absence of the use of the system} and \textit{the consequences of the use of the system for the rights and freedoms of all people concerned}. Finally, Article 5(3) states that in any case this type of deployment shall be subject to a prior authorisation by a judicial or other relevant authority. 
Although as mentioned before, there could be modifications regarding FR in the final text, the proposal mentions its proportional use by authorities in law enforcement scenarios. 

Another relevant initiative at the international level is led by the World Economy Forum (WEF), which  has recently developed a policy framework made of a set of principles 
for the use of FR in law enforcement~\cite{louradour2021policy}. 
The proposal identifies \emph{necessary and proportional use} as one of the principles to be followed, which is related to the trade-off between security threats and fundamental rights. It states that \emph{the decision to use facial recognition technology should always be guided by the objective of striking a fair balance between allowing law enforcement agencies to deploy the latest technologies, which are demonstrated to be accurate and safe, to safeguard individuals and society against security threats, and the necessity to protect the human rights of individuals. 
As a general principle, FR is considered to be linked to a cause and need as otherwise it would undermine human and fundamental rights.} 
This principle also refers to the need to document and justify the deployment of FR, specifying the classes of crimes or investigations for which its use is acceptable and/or lawful, and limiting the collection of images from public and publicly accessible spaces in terms of area and time period. 
In particular, it calls to consider alternatives to the use of FR and to ensure that its use is appropriate, limited and exclusively related to investigative purposes. 

Even though they are not policy initiatives, this section includes the efforts made by some private companies, research institutions and public sector organizations around the world to build ethical principles and guidelines for AI.
There is no consensus yet about the actual constituent elements of AI ethics, but the exhaustive analysis of 84 AI ethical principles/guidelines carried out by Jobin et al.~\cite{jobin2019global} finds that a global agreement is emerging around the following key principles: transparency, fairness, non-maleficence, responsibility, privacy, beneficence, freedom and autonomy and trust. 
These principles therefore apply to face recognition systems and are, indeed, aligned with both the aforementioned specific FR policies and citizens' concerns.

%% file: 3-2dplanes.tex
Two dimensional (2D) frameworks have been widely used as a simple but robust tools for resource allocation and policy intervention decisions in different fields as varied as Meteorology~\cite{Thompson:1952,wilks2001skill}, Economy~\cite{mishan2020cost} or Medicine~\cite{anderson1986policy,neumann2016cost}. 
In general, these frameworks compare costs vs benefits to assess the desirability of a project, a decision, or any other type of intervention. 
Although the words \textit{cost} and \textit{benefit} might sound purely economic, it should be noted that the trade-off between these two terms does not necessarily have to be monetary. 
For example, the cost of implementing an intervention or not can also be ethical (e.g., losing a fundamental right such as privacy) or medical (e.g., contracting a disease). Some 2D frameworks from other fields that have inspired this paper are described below.

Wilks proposes an economic cost/loss framework for weather forecast conceived for decision-makers~\cite{wilks2001skill}. On the one hand, this framework considers cost $C$ of implementing measures (i.e., to intervene) to protect against the effects of a potential severe weather condition and probability forecast $p$ that such event occurs. On the other hand, the occurrence of adverse weather events without this intervention would result in damage loss $L$. The intervention is considered to be economically viable when the cost/loss ratio is below the probability of occurrence of the adverse weather event, i.e., $\frac{C}{L} < p$. Thus, this framework transforms a  weather forecast into a GO/NO-GO decision. Also related to weather, Keith~\cite{keith2003optimization} proposes a flight deviation intervention model in the case of severe climate threats. The intervention based on an  adverse forecast involves loading additional fuel to reach an alternative airport. If no protection measures are taken and the event occurs, the flight should return to the airport of departure paying a higher cost in terms of fuel and delays.

The field of Medicine has been using Cost-Effectiveness Analysis (CEA) for a long time to decide whether intervene when confronted with a health threat. Decisions such as the allocation of extra health care resources~\cite{russell1996role} or population vaccination (e.g., for COVID-19~\cite{li2022cost}) are assessed.
Black~\cite{black1990plane} proposes a visual approach to CEA in Medicine by using a 2D plane, where the $x$-axis represents effectiveness (E), and the $y$-axis is cost (C). It defines a linear function with slope $K>0$, representing the maximum acceptable cost/effectiveness ratio, which splits the space into two regions. 
An intervention strategy is considered cost-effective if it provides more effectiveness than costs. 
Geometrically speaking, this implies that the 
point in the 2D plane representing the intervention is located at the bottom of the space where $E>\frac{C}{K}$. Two alternative intervention strategies, $I_1$ and $I_2$, can be evaluated in terms of their distance to the line with slope $K$, to decide which one is more cost-effective. 
It is important to highlight that, in the case of Medicine, the cost is economic, but the benefits are purely health-related (e.g., cost per COVID-19 contagion averted). 

\begin{figure}
\centering
  \includegraphics[width=0.99\columnwidth]{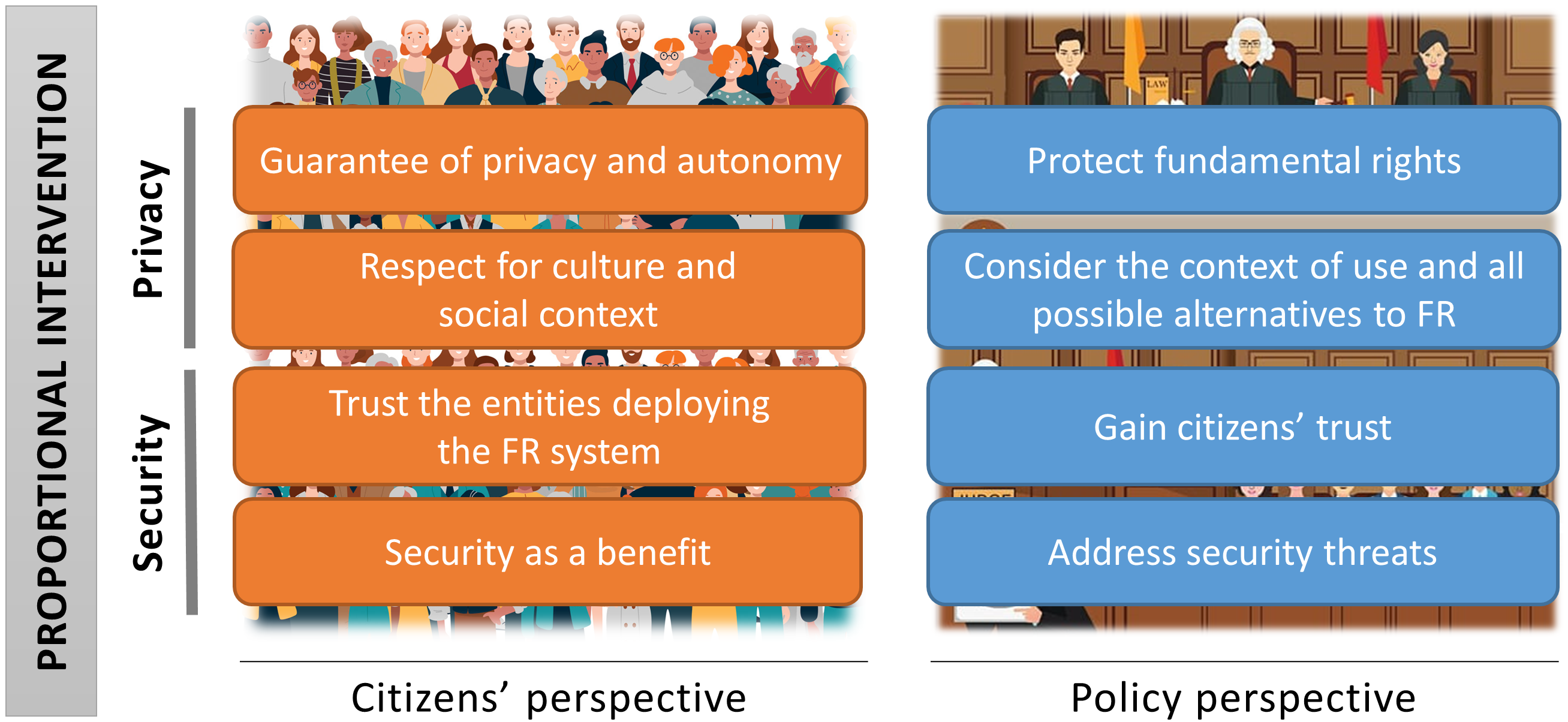}
  \caption{Key elements that a face recognition intervention decision must weigh, according to citizen and policy needs. }
  \label{fig:intervention_factors}
\end{figure}

This paper defines \textit{intervention} as the decision whether to deploy an FR system for law enforcement purposes as a protective action in the case of imminent public threat. While most papers focus on analyzing and improving the accuracy and performance of FR models, the assessment of real-world interventions has been widely ignored. 
As seen above, many factors might affect this decision.
Just like a severe weather threat case, FR watchlist suspects can cause varying levels of loss, damage, and harm affecting society from an economic and human life perspective. 
However, while severe weather cannot be stopped, a watchlist suspect can. 
Another factor to consider is the specific context in which the intervention would take place (e.g., in an indoor/open, more or less crowded space). 
In addition, FR deployment pays an ethical cost in terms of privacy-related fundamental rights~\cite{UN}. 
This represents a trade-off between ethical concerns and security needs, as represented in Figure~\ref{fig:intervention_factors}.

%% file: 4-Framework.tex

The cost/loss policy paradigm is used as a basis to propose a 2D graphical framework to assess the proportional and adequate use of FR in relevant contexts. 
It considers key factors, such as type of surveillance scenario, security risk, citizens' privacy concerns, and intends to assist authorities to arrive at an intervention decision.

The decision framework consists of two elements: a static 2D plane with \textit{Privacy Loss} vs \textit{Security Harm} variables, and a dynamic function $s_i$ driven by the implementation details.

\subsection{Proportional 2D-plane}
\label{sec:dimensions}
Our framework is based on a 2D cartesian plane modeling the proportional use of an FR intervention.
The y-axis considers the ethical cost of deployment, related to privacy and potential breaches of fundamental rights and associated with a loss of citizen trust in authorities. 
The x-axis represents the level of harm of the security threat, i.e., the harm potentially caused by not finding the individual on the watchlist (e.g., threat of not finding a criminal or a missing person). 
It should be noted that the economic dimension is intentionally not considered in our framework, as the focus is exclusively on the ethical aspects of FR interventions. 

As the ethical cost is strongly driven by invasion of privacy, the y-axis dimension has been named \textit{Privacy Loss}. 
Its $p$ value is formalized as mainly dependent on two variables: 

\begin{itemize}
    \item $d$ -- the density of people (e.g. people/hour) circulating in the deployment site and thus subject to FR. The higher the density of people under FR surveillance, the higher the overall loss of privacy. 
    \item $c$ -- the ethical cost linked to the site of deployment, which might be considered differently depending on its characteristics (e.g., public open space, indoor space, critical infrastructure), the intensity of surveillance (e.g., number of cameras in place, area covered) and the cultural context (e.g. benefit perceived by society in the country of deployment).
\end{itemize}

Then, the $p$ value of the \textit{Privacy Loss} dimension can be defined with the following function $f$: $p = f(d, c)$

The x-axis dimension has been named \textit{Security Harm}, representing the value of harm $h$, which could be mitigated by an FR deployment in site $i$. 
It covers both potential material and human harm with varying levels, from physical harm to human lives, and it depends on $d$ and variable $l$, representing the level of harm that the individual(s) being searched could potentially imply or cause.

Value $h$ of the \textit{Security Harm} dimension can then be defined with a function $g$ as follows: $h = g(d, l)$.

Table \ref{tab:cost} provides some examples of scenarios and how they may be linked to different $p$ and $h$ values respectively. 
It should be noted that, even though the framework's dimensions are conceived as continuous, conceptual values are provided ($p_n$ for \textit{Privacy Loss} and $h_m$ for \textit{Security Harm}) as their concrete numerical value might need to be adapted to the particular context of deployment including for instance  cultural considerations.

\input{tab/tab_2dplane.tex}


In the case of \textit{Privacy Loss} values $p_n$, index $n$ increases with the level of invasion of privacy. 
Its lowest value $p_1$  represents a scenario involving a small-to-medium-sized group of people captured by the FR system in outdoor spaces.
In the second privacy level, $p_2$, we consider scenarios involving a larger flow of people but this time in indoor spaces such as stadiums, airports or concerts.  
In this kind of spots, it is common for authorities to implement 
security measures at the entrance such as asking for IDs or tickets. 
Moreover, severe security incidents have recently ocurred in these scenarios, which has raised awareness and fear in the population~\cite{oksanen2020perceived}, consequently making them more open to FR intervention for the sake of security. 
The highest privacy level considered, $p_3$, is directly associated with FR intervention  
at so-called \textit{critical infrastructures} which include bus, metro or train stations~\cite{gritzalis2019critical} with very high circulation of people. 
Serious security incidents have also recently occurred in these scenarios. 
In these contexts, hundreds or even millions of people might be walking in front of the FR system everyday, unaware of the fact that their faces are being matched against those on a watchlist. 
As for \textit{Privacy Loss}, two levels are defined for \textit{Security Harm} $h_m$, where index $m$ increases with the severity of the harm that an individual on the watchlist might perpetrate. 
Table~\ref{tab:cost} describes the two proposed levels of \textit{Security Harm} $h_1$ and $h_2$, which are linked to human lives, distinguishing between murders/kidnapping/missing people and terrorist attacks, respectively. 
Note that in the case of kidnapping, the individual(s) being searched may be either the kidnapper(s) or the kidnapped person (or both). 
The rationale behind these harm levels is that the search for suspects of kidnapping, murders and terrorist attacks may be deemed sufficient to justify an FR intervention. 
Similarly, preventing any of these events, especially when there is a high probability of appearance $w$ of the searched person(s), may also be considered a justification for deployment as a protective action.
Note that the security issues involving material damage, such as robbery or property damage, are considered in this study as a non-proportional use of FR.

\begin{figure}[h]
\centering
  \includegraphics[width=0.75\columnwidth]{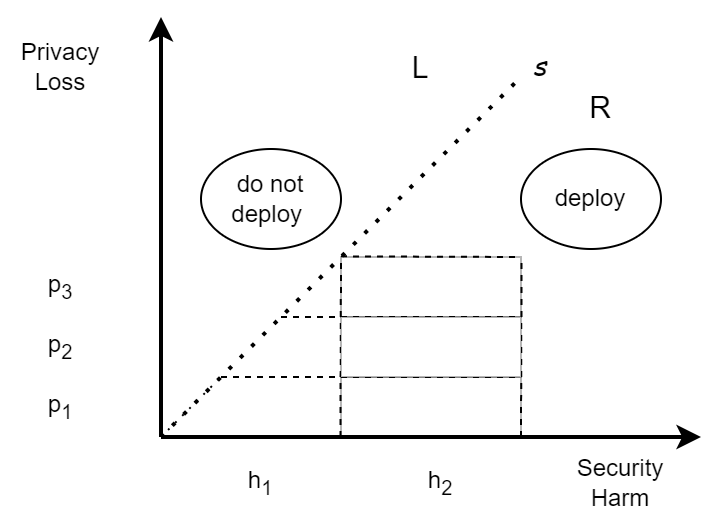}
  \caption{Main dimensions and visual elements of the  proportional 2D-plane proposed for FR intervention assessment.}
  \label{fig:regionI}
\end{figure}

Just like the cost/effectiveness 2D visualization proposed in~\cite{black1990plane} for the assessment of medical interventions, Fig.~\ref{fig:regionI} depicts our 2D plane which is divided into two regions by identity function $s$. 
In region R the value of \textit{Privacy Loss} $p$ is below that of \textit{Security Harm} $h$, and thus the use and deployment of FR may be deemed proportional. 
In region L, \textit{Privacy Loss} is above \textit{Security Harm} and FR deployment should be considered, in principle, non-proportional. 

Fig.~\ref{fig:regionI} also divides the discretized 2D space into rectangular regions or blocks based on the defined values of \textit{Privacy Loss} $p_n$  and \textit{Security Harm} $h_m$ in Table~\ref{tab:cost}.
The height/width (H/W) ratio of the blocks drives the graphical analysis of the \textit{Privacy Loss} vs \textit{Security Harm} trade-off proposed in this paper, which is further illustrated in the following sections.

Authorities facing this myriad of scenarios and having the responsibility to authorize an FR intervention would benefit from a decision framework helping them to weigh all these variables for citizen security. 
The proportional 2D-plane considers all those highly complex variables and provides a graphical and intuitive 2D representation to address the intervention decision. 

\subsection{The dynamic implementation function}\label{sec:function}

Fig.~\ref{fig:regionI} depicts identity line function $s$ dividing the 2D-plane into ''deploy'' and ''not-deploy'' regions.
In practice, the proposed framework uses a new dynamic implementation function, $s_i$, which will depend on the following variables: 
\begin{itemize}
    \item $w$ -- probability that the individual(s) on the watchlist  may appear in scenario $i$. This information might be provided, for example, by authorities or intelligence agencies based on previous investigations.
    \item $r$ -- FR system's reliability, for instance, in terms of false positives/negatives, false positive identification rate (FPIR) and demographic bias issues. For example, a false positive could result in the arrest of a wrong person and the consequent mistrust in the authorities. 
    \item $t$ -- period of time when the system is deployed (e.g., 24/7 in a venue, for a limited length of time during an event).
\end{itemize}

Thus, the dynamic function relates \textit{a priori} knowledge about the watchlist individual(s) in variable $w$, specification about the deployment in variable $t$, and details of the FR system in variable $r$.
This function is defined with the following equation:

\begin{equation}\label{eq:si}
s_i(h) = w \cdot h^{r} - t
\end{equation}

\noindent where $w$ is a probability variable defined in the range of $(0,1)$.
Variable $r$ is also defined in range $(0,1)$ and can be associated with the F1-score of the FR system. 
Finally, variable $t$ takes discretized values $[0, 0.25, 0.5]$ representing an FR deployment for a period of less than one week, a couple of weeks, and more than one month, respectively.

\begin{figure}[h]
\centering
  \includegraphics[width=0.75\columnwidth]{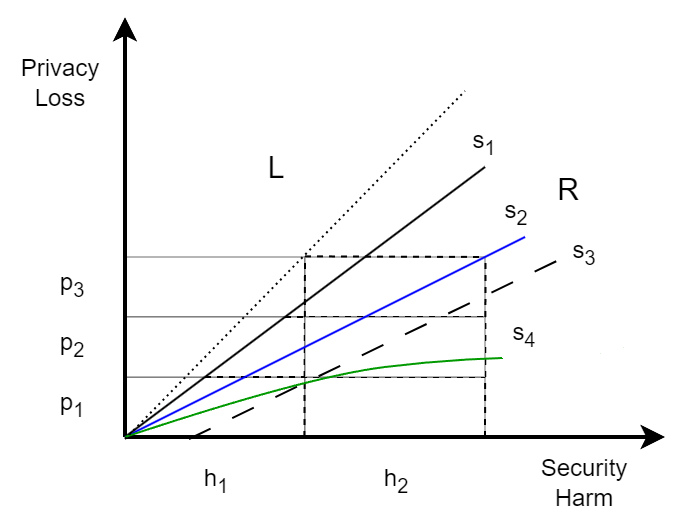}
  \caption{Examples of dynamic functions for different variables.}
  \label{fig:functionsSamples}
\end{figure}

Fig.~\ref{fig:functionsSamples} shows of different dynamic functions with different variable values.
Dynamic functions $s_1$ and $s_2$ have the following values $r=1$, $t=0$, and $w=0.75$ and $w=0.5$ respectively.
Function $s_3$ has these values $r=1$, $w=0.5$, and $t=0.25$, showing the same slope as $s_2$ with a rightward displacement, and the \textit{Privacy Loss} concern increases because of the longer deployment of the FR system.
Finally, $s_4$ is defined by $r=0.75$, $w=0.5$ and $t=0$.

For the sake of simplicity, in the following examples provided in this paper, variables $r=1$ and $t=0$ will be used, to work with straights lines without displacement.

\subsection{To deploy or not to deploy}

Both, the proportional 2D-plane (section~\ref{sec:dimensions}), and the dynamic function (section~\ref{sec:function}) determine the framework to address the intervention decision for a given law enforcement case.
Thus, the site of intervention and the watchlist individual(s) indicate the coordinates $(h_m,p_n)$ of the corresponding block in the 2D plane grid, as depicted in Fig.~\ref{fig:regionI}.
The specific variables of this case and the FR system shape the dynamic function, which will split the plane into a deployment region and a non-deployment region.
But instead of focusing on the entire R plane, the analysis is made at the block level.

\begin{figure}[h]
\centering
  \includegraphics[width=0.75\columnwidth]{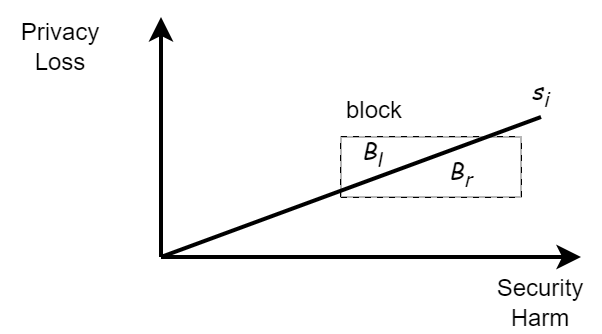}
  \caption{Block analysis based on surfaces $B_l$ and $B_r$ of the block at position $(h_m, p_n)$.}
  \label{fig:block_analisys}
\end{figure}

Fig.~\ref{fig:block_analisys} shows the graphical decision-making procedure.
Dynamic function $s_i$ splits block $B$ in two areas: $B_l$ and $B_r$.
Then, the 2D framework rule defines that the FR intervention on site $i$ is proportional if and only if $B_r > B_l$.

Now, we can return to Fig. \ref{fig:functionsSamples} to evaluate the proportional decision based on the different dynamic implementation functions.
Let us take block $(p_3, h_2)$. 
FR deployment with function $s_1$ determines an intervention decision, i.e., $B_r > B_l$, while function $s_2$ does not. 
As we know, the difference between both functions is appearance probability $w$.
A low $w$ at this high Privacy Loss level rules out the deployment decision.

Taking now block $(p_1, h_1)$, dynamic implementation function $s_2$ determines an intervention decision, but functions $s_3$ and $s_4$ do not.
This time, the difference for $s_3$ is a longer deployment time, and for $s_4$ it is a lower F1-score performance of the FR algorithm.

\subsection{Cultural Contexts}

\begin{figure*}[h]
\centering
\footnotesize
\begin{tabular}{ccc}
  \includegraphics[width=0.3\textwidth]{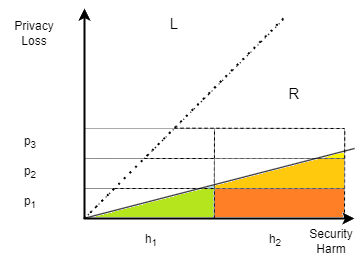} &
  \includegraphics[width=0.3\textwidth]{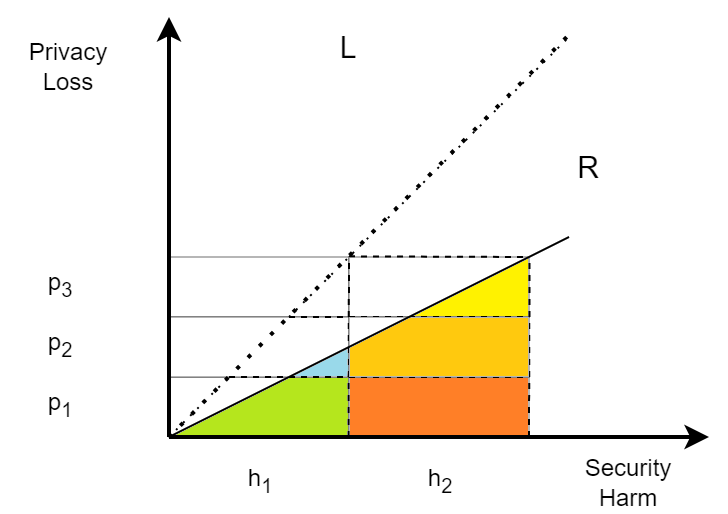} &
    \includegraphics[width=0.3\textwidth]{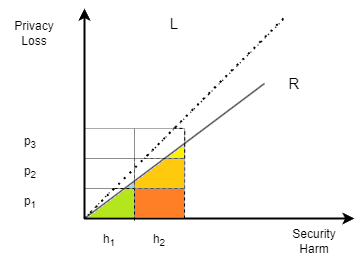} \\
    $s_i @ w = 0.25$ in tolerant context & $s_i @ w = 0.50$ in moderate context & $s_i @ w = 0.75$ in conservative context\\  
\end{tabular}
  \caption{Graphic illustration of how the proposed 2D plane would be used for the assessment of a FR intervention in sites $i$ with different associated $s_i$ dynamic function and H/W ratio.}  
  \label{fig:probaSamples2Dplane}
\end{figure*}

As mentioned before, public perception of facial processing applications on a wide range of scenarios concerning social good highly relates to cultural background~\cite{ritchie2021public,hupont2022landscape}.

Fig.~\ref{fig:probaSamples2Dplane} illustrates how the 2D plane would be used to drive an FR intervention decision in different cultural contexts.
From fig.~\ref{fig:block_analisys}, the intervention or non-intervention decision is based on the analysis of areas $B_l$ and $B_r$, showing the relevance of the height/width (H/W) ratio of the block.
Graphically, the FR deployment would be allowed when the colored filled area in a block is greater than the uncolored area, i.e., when $B_r > B_l$.
Fig.~\ref{fig:probaSamples2Dplane} depicts examples of different H/W ratios modeling different society perspectives about FR systems and Privacy Loss.
In the three examples, blocks $(p_1, h_1)$, $(p_1, h_2)$, $(p_2, h_2)$ fulfill the $B_r > B_l$ condition, and blocks $(p_2, h_1)$, $(p_3, h_2)$ do not follow the rule.
Values of $w$ leading to a similar intervention behavior are chosen, while the H/W ratio changes, with $r=1$ (meaning perfect FR performance), and $t=0$ (implying that the FR system will be deployed for a short time).

The first example (Fig.~\ref{fig:probaSamples2Dplane}-left), with  $H/W=\frac{3}{13}$, representing a context with high \textit{Security Harm} concerns, permits an FR intervention in block $(h_2, p_2)$ even with a low probability of subject appearance $w=0.25$, as shown by the corresponding dynamic function $s_i$.
This context accounts for a tolerant society towards FR systems and a moderate \textit{Privacy Loss} concern.
The second moderate context (Fig.~\ref{fig:probaSamples2Dplane}-center) has a ratio of $H/W=\frac{3}{9}$.
FR intervention in block $(h_2, p_2)$ would be deemed proportional when the probability of appearance is $w=0.5$ representing a reasonable value to deploy an FR system.
The last example (Figure~\ref{fig:probaSamples2Dplane}-left), where  $H/W=\frac{3}{5}$, accounts for a more conservative  policy context in terms of \textit{Privacy Loss} preservation.
In this case, FR deployment in block $(h_2, p_2)$ would only be worth with a very high probability of appearance of watchlist individual.
It means a dynamic function $s_i$ with $w=0.75$.

These examples demonstrate how cultural differences can drive the decision whether to intervene with FR or not, as well as the importance for policy makers to adequately address the \textit{Privacy Loss} vs \textit{Security Harm} trade-off.
Some countries and their citizens might be more willing than others to sacrifice part of their privacy in return for increased security.
However, thus far no concrete studies and figures on this matter have been developed.

%% file: tab/tab_2dplane.tex
\begin{table}[h]
    \centering
    \footnotesize
\begin{tabular}{p{0.7cm}p{0.4cm}p{6.1cm}}
\toprule
\multicolumn{3}{c}{\textit{Privacy Loss}} \\
\midrule
Privacy & Var & Description \\
   \cmidrule{1-3}
 $p_1$  & \textrm{d+} \textrm{c+}   &  FR deployed in a public open space with moderate people flow density (tens to hundreds of people per hour). Examples: streets, squares, neighbourhoods, etc.\\ 
   \cmidrule{1-3}
   $p_2$  & \textrm{d++} \textrm{c++}  & FR deployed in an indoor space with a moderate people flow density (hundreds of people passing by per hour) whith restricted access. Examples: airports or stadiums where people may enter with a ticket, such as a football match or musical concert.  
\\ 
   \cmidrule{1-3}
  $p_3$ & \textrm{d+++} \textrm{c+++}  &  FR deployed in a critical infrastructure with a high people flow density (circulation of hundreds to thousands of people per hour). This scenario could be, for instance, a mall, train, bus or metro station. \\ 
\midrule 
\multicolumn{3}{c}{\textit{Security Harm}} \\
\midrule 
Harm &  Var & Description \\
   \cmidrule{1-3}
 $h_1$  & \textrm{l++}  &  Security issues involving human life such as murder, kidnapping, or missing people.\\
   \cmidrule{1-3}
 $h_2$  &  \textrm{l+++} & Security issues concerning terrorists attacks linked to many human lives.\\
\bottomrule
\end{tabular}
    \caption{Scenarios examples for different levels of \textit{Privacy Loss} $p$ and \textit{Security Harm} $h$. The '+' sign indicates the variable's level of concern.}
    \label{tab:cost}
\end{table}

%% file: 5-SampleScenarios.tex
In the following section, the graphical 2D framework is applied to assess different types of FR interventions inspired by three real-world law enforcement scenarios. 

\subsubsection{Metropolitan Police Service Live Facial Recognition Trials}

\begin{figure}[h]
\centering
  \includegraphics[width=0.95\columnwidth]{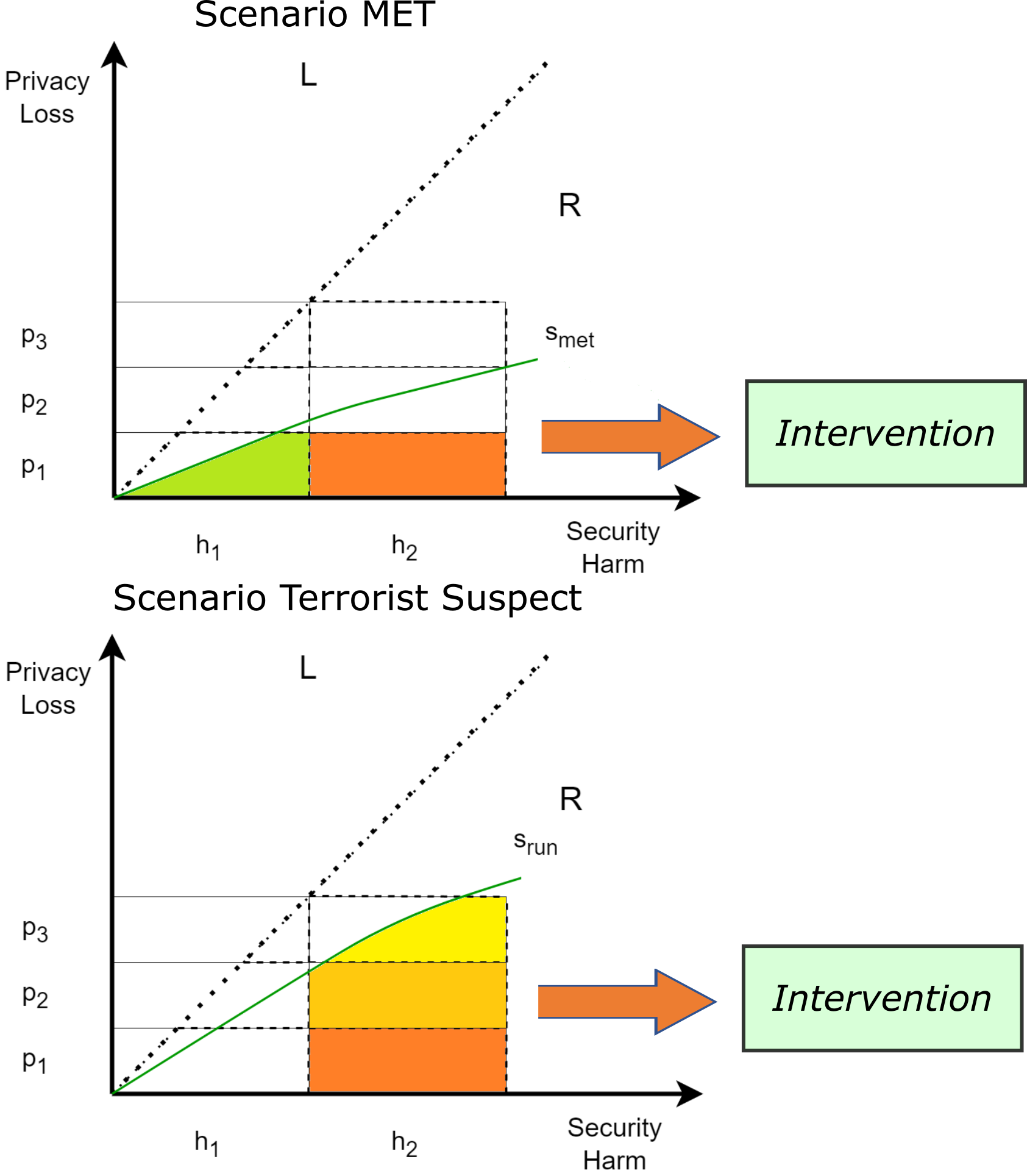}
  \caption{Framework in practice.}
  \label{fig:ejemplos}
\end{figure}

In 2020, London's Metropolitan Police Service presented a report about the deployment of Facial Recognition between August 2016 and February 2019\cite{MET}. 
The report details ten deployments in public spaces and a set of metrics of interest for a complete evaluation, including: duration, average of recognition opportunities, watchlist size, number of false alarms, number of people engaged by a police officer, and the number of actions/arrests.
Our focus will be on two of these trials, which used the same software version of the FR algorithm and similar equipment (surveillance camera).
Firstly, the “Stratford Westfield 28 June 2018” trial concerns a deployment on the street furniture for 6 hours.
The watchlist had 486 ‘Wanted Missing’ individuals chosen by geographic area (proximity to Westfield Stratford). 
The deployment produced 5 alerts over 10,000 detected and evaluated people.
It is worth mentioning here that the FR alerts (a match over a threshold) follow an operator adjudication process.
An operator is a qualified officer who has received advanced training in the facial recognition system and its features.
Thus, only one of these alerts resulted in engagement by an agent, but no action/arrest was performed.
Secondly, the “Romford February 2019” trial street deployment consisted of a 6:45 hour surveillance, detecting 10,100 pedestrians, with a larger watchlist of 1996 people including individuals wanted for violent offenses and filtered by geographic area. The deployment returned 13 positive matches, which on 3 occasions  led to an arrest.
In our framework, this \textit{Privacy Loss} scenario can be considered as $p_1$, given that it is deployed in a public open space with a moderate people flow density as defined in Table \ref{tab:cost}. 
The variables to define dynamic implementation function $s_{met}$ are: $t=0$ (limited time), $r=0.85$ (obtained from detailed performance statistics on the report, such as: False Alarms, and Positive Identifications at each trial), and probability $w=0.3$, which is a moderate value, because the watchlist is filtered by geographic area (individuals living in the watched neighborhood). The information about the level of harm associated with the individuals on the watchlist is missing.
This would allow authorities to determine the security harm value and, therefore, its proportionality, according to our model.
In fig. \ref{fig:ejemplos} this scenario is depicted, with $(p_1, h_1)$ and $(p_1, h_2)$ blocks colored under the dynamic function $s_{met}(h)$.
At this value of $w$, both areas below $s_{met}(h)$ indicate an \textit{Intervention} recommendation, but only if the \textit{Security Harm} caused by individuals on the watchlist corresponds to $h_1$ and $h_2$ levels.  

\subsubsection{Arrest of Terrorist Suspect in London}

A 21-year-old member of the British Army turned suspected terrorist and spy in January 2023 was sent to HMP Wandsworth prison. 
He escaped from the prison on the morning of Wednesday 6 September and was recaptured on Saturday 9 September.
The four-day search was coordinated at the Counter Terrorism Operations Centre (CTOC) in West Brompton, central London~\cite{KALIFE}. The situation room at the center had access to “cutting edge” spy technology including facial recognition, a CCTV camera network, and phone tracking data.
This case represents a real scenario with one individual on the watchlist of the FR system. 
Technical information about the deployment of the FR system is not available. 
However, this kind of search involves the use of FR in scenarios with different privacy losses: $p_1$, $p_2$, and $p_3$. 
While the level of harm cannot be exactly defined, the charges against the individual involve national security, and, thus, it can be assigned $h_3$. 
Other variables of the 2D framework can also be estimated to draw the dynamic implementation function $s_{run}$. 
The time parameter was less than one week, i.e., $t=0$. 
Variable $r$ can be considered as $r=0.9$. 
Finally, if it is considered that the FR deployment is performed in places where the public reported sighting of the suspect, appearance probability is $w=0.75$. 
Also, scenarios involving $p_3$ typically correspond to those places where a fugitive in the run can show up and escape using subways, trains, or plains.
In this case, deployment could be considered within the area of proportional use, if it is associated with a high potential harm, which should be regarded with respect to a maximum privacy threshold.
Fig. \ref{fig:ejemplos} shows blocks $(p_1, h_2)$, $(p_2, h_2)$ and $(p_3, h_2)$ colored under the dynamic function validating   the \textit{Intervention} recommendation.

\subsubsection{Br\o ndby IF’s STADIUM}

Br\o ndby IF is a professional football club in the Danish Superliga \cite{Stadium}. 
In the summer of 2019 at Br\o ndby IF’s stadium, Panasonic installed “FacePRO,” a facial recognition system. Individuals who have been caught breaking stadium rules are banned from coming back to games and are registered on a watchlist. 
Br\o ndby IF has an average home game attendance of roughly 14,000 people, and approximately up to 100 people are registered on the watchlist on average.
For the graphical 2D framework, this scenario would correspond to a $p_2$ level of privacy, given that it is deployed in an indoor space as specified in Table \ref{tab:cost}. 
The time variable is $t=0$ because the deployment corresponds to a short period (the match time). 
The appearance probability has a relatively high value, i.e., $w=0.3$, because the fans are likely to be present at the match. 
The technology on the FR model is based on \cite{xiong2018good}, and the evaluation point from the official evaluation report of the National Institute of Standards and Technology (NIST) states $r=0.95$. 
However, when the level of security harm caused by individuals on the watchlist, cannot be determined from Table \ref{tab:cost}, as the subjects are banned for violent behavior not for severe crimes. 
Our 2D graphical framework does not place the scenario in the \textit{Intervention-Non Intervention} plane, which means that the FR deployment is not recommended. 
Thus, other  types of interventions that can be envisaged from Fig. \ref{fig:intervention_levels}.

%% file: 6-AI-ACT.tex
The European Union's Artificial Intelligence (AI) Act was published in the Official Journal on 12 July 2024~\cite{AIact}. 
The AI Act represents the inaugural globally binding horizontal regulation on AI, establishing a unified framework for the utilisation and provision of AI systems within the European Union (EU). 

It provides a classification system for AI systems, with differentiated requirements and obligations based on a risk-based approach. 
The utilisation of AI systems that present an \textit{unacceptable} level of risk is prohibited. A broad category of \textit{high-risk} AI systems, which have the potential to cause significant harm to individuals in terms of their health, safety or fundamental rights, is permitted to enter the EU market, but only subject to a set of requirements and obligations. 
Those AI systems that present only \textit{minimal risk} to people will not be subject to further obligations, whereas those AI systems that are opaque and therefore pose limited risks will be subject to information and transparency requirements. 

This section analyzes the proposed deployment framework which tackles most of the concerns of the AI Act. 

\subsection{AI Act Context}

The AI Act establishes a definitive stance on the deployment of AI systems in urban surveillance, emphasising that their utilisation for \textit{real-time} remote biometric identification of individuals in publicly accessible locations for the purpose of law enforcement is particularly invasive to the rights and freedoms of those concerned. 
This may have a significant impact on the private lives of a considerable proportion of the population, engendering a sense of constant surveillance and indirectly deterring the exercise of fundamental rights, including the freedom of assembly.

The use of those systems for the purpose of law enforcement should therefore be prohibited. 
However, there are important exceptions and guidelines for special scenarios:

\begin{itemize}
    \item Proportional Deployment: there exists an exhaustively listed and narrowly defined situations, where the use is strictly necessary to achieve a substantial public interest, the importance of which outweighs the risks.
    \item Scope Limitations: the biometric system should be employed solely for the purpose of confirming the identity of the individual whose identity is specifically targeted. The deployment of such systems should be limited to what is strictly necessary in terms of the period of time, as well as the geographic and personal scope.
    \item Authorisation: The deployment of remote biometric identification system should be contingent upon the prior issuance of an express and specific authorisation by a judicial authority or by an independent administrative authority of a Member State whose decision is binding. Such authorisation should, in principle, be obtained prior to the use of the AI system with a view to identifying a person or persons.
    \item Risk Assessment: The utilisation of the system in question should be authorised exclusively in instances where the pertinent law enforcement authority has conducted a fundamental rights impact assessment and has registered the system in the database in accordance with the stipulations set forth in AI Act regulation.
\end{itemize}

\subsection{Permitted AI practices}
Chapter II develops Prohibited AI practices in article 5 for several applications of such systems.
In regard to the implementation of real-time remote biometric identification systems, the following exceptions to their usage in point \textit{h} shall be observed, provided that such usage is deemed strictly necessary for the fulfilment of one or more of the following objectives:
\begin{itemize}
    \item[i] the targeted search for specific victims of abduction, trafficking in human beings or sexual exploitation of human beings, as well as the search for missing persons;
    \item[ii] the prevention of a specific, substantial and imminent threat to the life or physical safety of natural persons or a genuine and present or genuine and foreseeable threat of a terrorist attack;
    \item[iii] the localisation or identification of a person suspected of having committed a criminal offence, for the purpose of conducting a criminal investigation or prosecution or executing a criminal penalty for offences referred to in Annex II and punishable in the Member State concerned by a custodial sentence or a detention order for a maximum period of at least four years.
\end{itemize}

Point 2 of this article states that the use of such systems shall take into account the following elements:
\begin{itemize}
    \item[a] the nature of the situation giving rise to the possible use, in particular the seriousness, probability and scale of the harm that would be caused if the system were not used;
    \item[b] the consequences of the use of the system for the rights and freedoms of all persons concerned, in particular the seriousness, probability and scale of those consequences.
\end{itemize}

\subsection{Framework compliance to AI Act}

The preceding sections have provided an overview of the principles governing the utilization of real-time biometric identification systems. It is our contention that our framework encompasses the majority of the principal concepts pertaining to the decision-making process surrounding the deployment of such systems.
\begin{itemize}
    \item Scope: temporal and situations concepts are targeted in eq.~\ref{eq:si}, by considering the period of time of the deployment as well as the probability that the subject could be present at that location.
    \item Scale: level of harm and ethical costs are evaluated of the two dimensional plane and the discretized space.  
\end{itemize}

The framework fulfills the necessity of the authorities to evaluate the impact assessment of the use of the biometric system. Furthermore, it synthesizes a more detailed evaluation about the specific situation conducted by the policy forces, which could serve as the foundation for a decision by the judicial authority regarding the authorization or prohibition of the deployment of the system.

%% file: 7-Conclusions.tex
A 2D graphical framework was proposed to assess the proportional use of FR systems in real-world scenarios, grounded on an ethical cost vs security gain model. 
The two dimensions consider variables from recent studies and policies on face recognition and related citizen privacy concerns.
To the best of our knowledge, this is the first framework addressing the problem of FR intervention, which might have a high impact for decision makers and lead to new research considering the principle of proportionality in FR.
It will also hopefully contribute to open discussion, in line with worldwide regulations such as the European AI Act~\cite{AIact}, on the proportional and strictly necessary use of FR technology. 

Our framework has, however, some limitations. 
In its practical implementation, a simple linear approach has been used with a broad discretization of the 2D plane into large intervention blocks.
This model can be improved by incorporating in its design stakeholders directly involved in FR deployment (e.g., citizens, decision makers, etc.). 
To address this, future work may include developing simulations of different FR scenarios and conducting a large-scale user survey to understand which of them are deemed proportional as well as culture-related information. This will allow us to come up with a more fine-grained mathematical model taking advantage of the continuous nature of the variables, such as the H/W ratio.
Indeed, the framework needs to identify different cultural and ethical preferences by countries or world regions.
Future work should also test the usability and usefulness of the framework with policy-makers and authorities, and apply that kind of framework to other decision-making situations.